\pdfoutput=1

\documentclass[conference]{IEEEtran}

\ifCLASSINFOpdf
\else
\fi

\usepackage{amsmath,amsthm}
\usepackage{amssymb, bm}
\usepackage{float}
\usepackage{amsfonts}
\usepackage{graphicx}
\usepackage{color}
\usepackage{tikz}
\usepackage{tikzscale}
\usepackage{pgfplots}
\usepackage{graphicx}
\usepackage[nolist]{acronym} 
\pgfplotsset{compat=newest}
\usepackage{subcaption}
\usepackage{marginnote}

\usepackage{units}
\usepackage{amsmath, amsbsy, amssymb}

\usetikzlibrary{arrows}
\usetikzlibrary{shapes}
\usetikzlibrary{snakes}
\usetikzlibrary{chains}
\usetikzlibrary{positioning}
\usetikzlibrary{patterns}
\usetikzlibrary{calc}
\usetikzlibrary{fit}
\usetikzlibrary{dsp}
\usetikzlibrary{spy,backgrounds}

\definecolor{mittelblau}{RGB}{0, 126, 198}
\definecolor{violettblau}{cmyk}{0.9, 0.6, 0, 0}
\definecolor{rot}{RGB}{238, 28 35}
\definecolor{apfelgruen}{RGB}{140, 198, 62}
\definecolor{gelb}{RGB}{1, 221, 0}
\definecolor{orange}{RGB}{244, 111, 33}
\definecolor{pink}{RGB}{237, 0, 140}
\definecolor{lila}{RGB}{128, 10, 145}
\definecolor{hellgrau}{RGB}{224, 224, 224}
\definecolor{mittelgrau}{RGB}{128, 128, 128}
\definecolor{dunkelgrau}{RGB}{80,80,80}
\definecolor{anthrazit}{RGB}{19, 31, 31}

\newcommand{\disable}[1]{}
\newcommand{\sebastian}[1]{\textcolor{rot}{#1}}

\begin{document}

%
\title{Scaling Deep Learning-based Decoding of Polar Codes via Partitioning}

\author{\IEEEauthorblockN{Sebastian Cammerer\IEEEauthorrefmark{1}, Tobias Gruber\IEEEauthorrefmark{1}, Jakob Hoydis\IEEEauthorrefmark{2}, and Stephan ten Brink\IEEEauthorrefmark{1} }
\IEEEauthorblockA{
\IEEEauthorrefmark{1} Institute of Telecommunications, Pfaffenwaldring 47, University of  Stuttgart, 70659 Stuttgart, Germany 
    \\\{cammerer,gruber,tenbrink\}@inue.uni-stuttgart.de
\\
\IEEEauthorrefmark{2}Nokia Bell Labs, Route de Villejust, 91620 Nozay, France
    \\jakob.hoydis@nokia-bell-labs.com
}
}

\maketitle

\begin{acronym}
 \acro{NN}{neural network}
 \acro{NE}{normalized error}
 \acro{ECC}{error-correcting code}
 \acro{MLD}{maximum likelihood decoding}
 \acro{HDD}{hard decision decoding}
 \acro{SDD}{soft decision decoding}
 \acro{NND}{neural network decoding}
 \acro{GPU}{graphical processing unit}
 \acro{BP}{belief propagation}
 \acro{LDPC}{low density parity check}
 \acro{BER}{bit error rate}
 \acro{SNR}{signal to noise ratio}
 \acro{ReLU}{rectified linear unit}
 \acro{BPSK}{binary phase shift keying}
 \acro{AWGN}{additive white Gaussian noise}
 \acro{MSE}{mean squared error}
 \acro{LLR}{log likelihood ratio}
 \acro{MAP}{maximum a posteriori}
 \acro{NVE}{normalized validation error}
 \acro{BCE}{binary crossentropy}
 \acro{BLER}{block error rate}
 \acro{SC}{successive cancellation}
 \acro{SCL}{successive cancellation list}
 \acro{CRC}{cyclic redundancy check}
 \acro{FPGA}{field programmable gate array}
 \acro{ASIC}{application-specific integrated circuit}
 \acro{ML}{machine learning}
 \acro{PE}{processing element}
 \acro{PNN}{partitioned neural network}
  \acro{NVE}{normalized validation error}
  \acro{NN}{neural network}
  \acro{SSC}{simplified successive cancellation}
  \acro{SPC}{single-parity checks}
  \acro{RC}{repetition codes}
  \acro{PSCL}{partitioned successive cancellation list}
  \acro{IoT}{internet of things}
  \acro{HDPC}{high density parity check}
\end{acronym}

\begin{abstract}
The training complexity of deep learning-based channel decoders scales exponentially with the codebook size and therefore with the number of information bits. Thus, \ac{NND} is currently only feasible for very short block lengths. In this work, we show that the conventional iterative decoding algorithm for polar codes can be enhanced when sub-blocks of the  decoder are replaced by \ac{NN} based components. Thus, we partition the encoding graph into smaller sub-blocks and train them individually, closely approaching \ac{MAP} performance per sub-block. These blocks are then connected via the remaining conventional belief propagation decoding stage(s). The resulting decoding algorithm is non-iterative and inherently enables a high-level of parallelization, while showing a competitive \ac{BER} performance. We examine the degradation through partitioning and compare the resulting decoder to state-of-the-art polar decoders such as successive cancellation list and belief propagation decoding.
\end{abstract}

\IEEEpeerreviewmaketitle

\acresetall

\section{Introduction}

Non-iterative and consequently low-latency decoding together with close to \ac{MAP} decoding performance are two advantages  of deep learning-based channel decoding.
However, this concept is mainly restricted by its limited scalability in terms of the supported block lengths, known as \textit{curse of dimensionality} \cite{wang1996}. For $k$ information bits, the \ac{NN} needs to distinguish between $2^k$ different codewords, which results in an exponential training complexity in case that the full codebook needs to be learned. In this work we focus on \ac{NN} decoding of polar codes and show that the scalability can be significantly improved towards practical lengths when the \ac{NN} only replaces sub-components of the current decoding algorithm.

List decoding \cite{talvardyList} with manageable list sizes and excellent \ac{BER} performance for short block lengths makes polar codes \cite{ArikanMain} a potential candidate for future communication standards such as the upcoming 5G standard or \emph{\ac{IoT}} applications. As polar codes are currently proposed for the 5G control channel \cite{RAN87}, decoding algorithms for very short block lengths are of practical importance \cite{liva2016}. Besides that, the rate can be completely flexible adjusted with a single-bit granularity. However, the price to pay is an inherently serial decoding algorithm which is in general hard to accelerate, e.g., through parallel processing \cite{cammerer_HybridGPU}. This leads to high decoding latency when compared to state-of-the-art \ac{LDPC} codes/decoders \cite{giard_hardwareDecoders}. Thus, there is a demand for alternative decoding strategies. Besides algorithmic optimizations \cite{giard_hardwareDecoders} of the existing algorithms, modifying the code structure \cite{siegel_enhanced_bp} can be considered to overcome this issue. However, once standardized, the encoder cannot be changed. In this work, we propose an alternative approach by applying machine learning techniques to find an alternative decoding algorithm instead of changing the code structure. Once trained, the final decoding algorithm (i.e., the weights of a deep neural network) itself is static and can be efficiently implemented and parallelized on a \ac{GPU}, \ac{FPGA}, or \ac{ASIC}.

A first investigation of the topic \emph{learning to decode} was already done in \cite{wang1996}. The authors showed that the main difficulty lies within the \emph{curse of dimensionality} meaning that for $k$ information bits $2^k$ classes exist, leading to exponential complexity during the training phase. In other applications, such as computer vision, the number of possible output classes is typically limited, e.g., to the number of different objects. In contrast to many other machine learning fields, an unlimited amount of labeled training data is available, since the encoding function and the channel model are well known. Additionally, a clear benchmark with existing decoders is possible.
Although very powerful machine learning libraries such as Theano \cite{alrfou_theano} and Tensorflow \cite{abadi_tensorflow} are available nowadays and the computation power increased by order of magnitudes, the exponential complexity still hinders straight-forward learning of practical code lengths as shown in \cite{gruber_CISS}. It was observed in \cite{gruber_CISS}, that there is a certain generalization of \ac{NN} decoding, meaning that the \ac{NN} can infer from certain codewords to others it has never seen before. This is essential for learning longer codes and gives hope that \ac{NND} can be scaled to longer codes. However, to the best of our knowledge, the naive approach of learning to decode only works for rather small block lengths.

The authors in \cite{nachmani_MLDecoding} proposed the idea of using machine learning techniques to train the weights of a belief propagation factor graph in order to improve its decoding performance for \ac{HDPC} codes. As the Tanner graph is already given initially and only its weights are refined, their approach scales very well for larger block lengths and does not suffer from the curse of dimensionality. However, in this case, the use of machine learning refines an existing solution. The decoding algorithm itself is not learned, since the iterative nature of the BP algorithm is kept.

In our work, we tackle the problem of completely replacing the polar code decoder by a machine learning approach. As it turns out, only small codeword lengths can be trained efficiently, and thus we divide the polar encoding graph into several sub-graphs (cf. \cite{hashemi_partitioned}). We learn sub-block wise decoding and couple the components by conventional \ac{BP} stages. This scales the results from \cite{gruber_CISS} towards practical block lengths.

\section{Polar Codes}

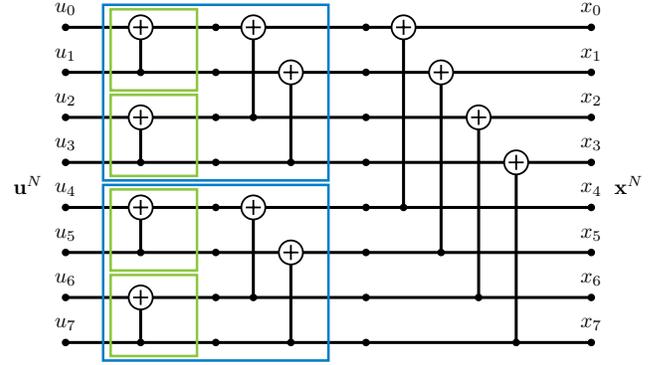
\begin{figure}[t]
\vspace*{0.3cm}
\centering
\resizebox{\columnwidth}{!}{
  \begin{tikzpicture}[scale=1,y=0.75cm,x=1.25cm]
\node[dspnodefull,minimum size=1mm] (u1) at (0, 1) {$u_6$};
\node[dspnodefull,minimum size=1mm] (u2) at (0, 0) {$u_7$};
\node[dspnodefull,minimum size=1mm] (temp) at (1, 0) {};
\node[dspadder](xor) at (1,1) {};
\node[dspnodefull,minimum size=1mm] (x1) at (2, 1) {};
\node[dspnodefull,minimum size=1mm] (x2) at (2, 0) {};
\draw[line width = 0.5mm](u1)--node[above] {}(xor);
\draw[line width = 0.5mm](temp)--(xor);
\draw[line width = 0.5mm](xor)--node[above] {}(x1);
\draw[line width = 0.5mm](temp)--node[above] {}(x2);
\draw[line width = 0.5mm](u2)--node[above] {}(temp);
\node[dspnodefull,minimum size=1mm] (u12) at (0, 3) {$u_4$};
\node[dspnodefull,minimum size=1mm] (u22) at (0, 2) {$u_5$};
\node[dspnodefull,minimum size=1mm] (temp2) at (1, 2) {};
\node[dspadder](xor2) at (1,3) {};
\node[dspnodefull,minimum size=1mm] (x12) at (2, 3) {};
\node[dspnodefull,minimum size=1mm] (x22) at (2, 2) {};
\draw[line width = 0.5mm](u12)--node[above] {}(xor2);
\draw[line width = 0.5mm](temp2)--(xor2);
\draw[line width = 0.5mm](xor2)--node[above] {}(x12);
\draw[line width = 0.5mm](temp2)--node[above] {}(x22);
\draw[line width = 0.5mm](u22)--node[above] {}(temp2);
\node[] (u13) at (-0.5,3.5) {$\mathbf{u}^N$};
\node[] (u13) at (7.5,3.5) {$\mathbf{x}^N$};
\node[dspnodefull,minimum size=1mm] (u13) at (0, 5) {$u_2$};
\node[dspnodefull,minimum size=1mm] (u23) at (0, 4) {$u_3$};
\node[dspnodefull,minimum size=1mm] (temp3) at (1, 4) {};
\node[dspadder](xor3) at (1,5) {};
\node[dspnodefull,minimum size=1mm] (x13) at (2, 5) {};
\node[dspnodefull,minimum size=1mm] (x23) at (2, 4) {};
\draw[line width = 0.5mm](u13)--node[above] {}(xor3);
\draw[line width = 0.5mm](temp3)--(xor3);
\draw[line width = 0.5mm](xor3)--node[above] {}(x13);
\draw[line width = 0.5mm](temp3)--node[above] {}(x23);
\draw[line width = 0.5mm](u23)--node[above] {}(temp3);
\node[dspnodefull,minimum size=1mm] (u14) at (0, 7) {$u_0$};
\node[dspnodefull,minimum size=1mm] (u24) at (0, 6) {$u_1$};
\node[dspnodefull,minimum size=1mm] (temp4) at (1, 6) {};
\node[dspadder](xor4) at (1,7) {};
\node[dspnodefull,minimum size=1mm] (x14) at (2, 7) {};
\node[dspnodefull,minimum size=1mm] (x24) at (2, 6) {};
\draw[line width = 0.5mm](u14)--node[above] {}(xor4);
\draw[line width = 0.5mm](temp4)--(xor4);
\draw[line width = 0.5mm](xor4)--node[above] {}(x14);
\draw[line width = 0.5mm](temp4)--node[above] {}(x24);
\draw[line width = 0.5mm](u24)--node[above] {}(temp4);
\node[dspnodefull,minimum size=1mm] (temp5) at (3, 0) {};
\node[dspadder](xor5) at (3,2) {};
\node[dspnodefull,minimum size=1mm] (y22) at (4, 0) {};
\node[dspnodefull,minimum size=1mm] (y42) at (4, 2) {};
\draw[line width = 0.5mm](x22)--node[above] {}(xor5);
\draw[line width = 0.5mm](temp5)--(xor5);
\draw[line width = 0.5mm](xor5)--node[above] {}(y42);
\draw[line width = 0.5mm](temp5)--node[above] {}(y22);
\draw[line width = 0.5mm](x2)--node[above] {}(temp5);
\node[dspnodefull,minimum size=1mm] (temp6) at (2.5, 1) {};
\node[dspadder](xor6) at (2.5,3) {};
\node[dspnodefull,minimum size=1mm] (y12) at (4, 1) {};
\node[dspnodefull,minimum size=1mm] (y32) at (4, 3) {};
\draw[line width = 0.5mm](x12)--node[above] {}(xor6);
\draw[line width = 0.5mm](temp6)--(xor6);
\draw[line width = 0.5mm](xor6)--node[above] {}(y32);
\draw[line width = 0.5mm](temp6)--node[above] {}(y12);
\draw[line width = 0.5mm](x1)--node[above] {}(temp6);
\node[dspnodefull,minimum size=1mm] (temp7) at (3, 4) {};
\node[dspadder](xor7) at (3,6) {};
\node[dspnodefull,minimum size=1mm] (y6) at (4, 4) {};
\node[dspnodefull,minimum size=1mm] (y8) at (4, 6) {};
\draw[line width = 0.5mm](x24)--node[above] {}(xor7);
\draw[line width = 0.5mm](temp7)--(xor7);
\draw[line width = 0.5mm](xor7)--node[above] {}(y8);
\draw[line width = 0.5mm](temp7)--node[above] {}(y6);
\draw[line width = 0.5mm](x23)--node[above] {}(temp7);
\node[dspnodefull,minimum size=1mm] (temp8) at (2.5, 5) {};
\node[dspadder](xor8) at (2.5,7) {};
\node[dspnodefull,minimum size=1mm] (y5) at (4, 5) {};
\node[dspnodefull,minimum size=1mm] (y7) at (4, 7) {};
\draw[line width = 0.5mm](x14)--node[above] {}(xor8);
\draw[line width = 0.5mm](temp8)--(xor8);
\draw[line width = 0.5mm](xor8)--node[above] {}(y7);
\draw[line width = 0.5mm](temp8)--node[above] {}(y5);
\draw[line width = 0.5mm](x13)--node[above] {}(temp8);
\node[dspnodefull,minimum size=1mm] (temp9) at (4.5, 3) {};
\node[dspadder](xor9) at (4.5,7) {};
\node[dspnodefull,minimum size=1mm] (z1) at (7, 3) {$x_4$};
\node[dspnodefull,minimum size=1mm] (z5) at (7, 7) {$x_0$};
\draw[line width = 0.5mm](y7)--node[above] {}(xor9);
\draw[line width = 0.5mm](temp9)--(xor9);
\draw[line width = 0.5mm](xor9)--node[above] {}(z5);
\draw[line width = 0.5mm](temp9)--node[above] {}(z1);
\draw[line width = 0.5mm](y32)--node[above] {}(temp9);
\node[dspnodefull,minimum size=1mm] (temp10) at (5, 2) {};
\node[dspadder](xor10) at (5,6) {};
\node[dspnodefull,minimum size=1mm] (z2) at (7, 2) {$x_5$};
\node[dspnodefull,minimum size=1mm] (z6) at (7, 6) {$x_1$};
\draw[line width = 0.5mm](y8)--node[above] {}(xor10);
\draw[line width = 0.5mm](temp10)--(xor10);
\draw[line width = 0.5mm](xor10)--node[above] {}(z6);
\draw[line width = 0.5mm](temp10)--node[above] {}(z2);
\draw[line width = 0.5mm](y42)--node[above] {}(temp10);
\node[dspnodefull,minimum size=1mm] (temp11) at (5.5, 1) {};
\node[dspadder](xor11) at (5.5,5) {};
\node[dspnodefull,minimum size=1mm] (z3) at (7, 1) {$x_6$};
\node[dspnodefull,minimum size=1mm] (z7) at (7, 5) {$x_2$};
\draw[line width = 0.5mm](y5)--node[above] {}(xor11);
\draw[line width = 0.5mm](temp11)--(xor11);
\draw[line width = 0.5mm](xor11)--node[above] {}(z7);
\draw[line width = 0.5mm](temp11)--node[above] {}(z3);
\draw[line width = 0.5mm](y12)--node[above] {}(temp11);
\node[dspnodefull,minimum size=1mm] (temp12) at (6, 0) {};
\node[dspadder](xor12) at (6,4) {};
\node[dspnodefull,minimum size=1mm] (z4) at (7, 0) {$x_7$};
\node[dspnodefull,minimum size=1mm] (z8) at (7, 4) {$x_3$};
\draw[line width = 0.5mm](y6)--node[above] {}(xor12);
\draw[line width = 0.5mm](temp12)--(xor12);
\draw[line width = 0.5mm](xor12)--node[above] {}(z8);
\draw[line width = 0.5mm](temp12)--node[above] {}(z4);
\draw[line width = 0.5mm](y22)--node[above] {}(temp12);

\draw[color=mittelblau,very thick] (0.5,-0.4) rectangle (3.5,3.5);
\draw[color=mittelblau,very thick] (0.5,3.6) rectangle (3.5,7.5);

\draw[color=apfelgruen,very thick] (0.6,-0.3) rectangle (1.75,1.5);
\draw[color=apfelgruen,very thick] (0.6,1.6) rectangle (1.75,3.4);
\draw[color=apfelgruen,very thick] (0.6,3.7) rectangle (1.75,5.5);
\draw[color=apfelgruen,very thick] (0.6,5.6) rectangle (1.75,7.4);

\end{tikzpicture}
 }
\caption{Polar Encoding circuit for $N=8$; blue boxes indicate the independent partitions of the code for $M=2$ and green boxes for $M=4$. }
\label{encoding-circuit}
\end{figure}

An encoder for polar codes maps the $k$ information bits onto the $k$ most reliable bit positions of the vector $\mathbf{u}$ of length $N$, denoted as information set $\mathbb{A}$, while the remaining $N-k$ positions are treated as frozen positions. These frozen positions are denoted as $\mathbb{\bar{A}}$ and must be known at the decoder side. Now the input block $\mathbf{u}$ is encoded according to $\mathbf{x}=\mathbf{u}\cdot\mathbf{\mathbf{G}}_{N}$, where ${\mathbf{G}=\mathbf{F}^{\otimes n}}$ is the generator matrix and $\mathbf{F}^{\otimes n}$ denotes the $n^{th}$ Kronecker power of the kernel $\ensuremath{\mathbf{F}=\left[\begin{smallmatrix}1 & 0\\1 & 1\end{smallmatrix}\right]}$. The resulting encoding circuit is depicted in Fig. \ref{encoding-circuit} for ${N=8}$, which also defines the decoding graph. This factor graph consists of ${n+1=\log_{2}(N)+1}$ stages, each consisting of $N$ nodes. 

The \ac{BER} performance of a polar code highly depends on the type of decoder used and has been one of the most exciting and active areas of research related to polar coding. There are two main algorithmic avenues to tackle the polar decoding problem:
\begin{enumerate}
\item \acl{SC}-based decoding, following a serial ``channel layer unwrapping'' decoding strategy \cite{ArikanMain},
\item \acl{BP}-based decoding based on Gallager's \ac{BP} iterative algorithm \cite{gallager_ldpc}.
\end{enumerate}
Throughout this work, we stick with the \ac{BP} decoder as its structure is a better match to neural networks and enables parallel processing. For details about \ac{SC} decoding and its list extension called \ac{SCL} decoding, we refer the interested reader to \cite{talvardyList} and  \cite{ArikanMain}. 
The \ac{BP} decoder describes an iterative message passing algorithm with soft-values, i.e., \ac{LLR} values over the encoding graph.
For the sake of simplicity, we assume \ac{BPSK} modulation and an \ac{AWGN} channel. However, other channels can be implemented straightforwardly. For a received value $y$, it holds that
$$LLR(y)=\operatorname{ln}\left(\frac{P(x=0|y)}{P(x=1|y)}\right)=\frac{2}{\sigma^2}y$$  where $\sigma^2$ is the noise variance.
There are two types of \ac{LLR} messages: the right-to-left messages ($\mathbf{L}$-messages) and the left-to-right messages ($\mathbf{R}$-messages). One \ac{BP} iteration consists of two update propagations:
\begin{enumerate}
\item Left-to-right propagation: The $\mathbf{R}$-messages are updated starting from the leftmost stage (i.e., the stage of a priori information) until reaching the rightmost stage. 
\item Right-to-left propagation: The $\mathbf{L}$-messages are updated starting from the rightmost stage (i.e., the stage of channel information) until reaching the leftmost stage.
\end{enumerate}
The output from two nodes becomes the input to a specific neighboring \ac{PE} (for more details we refer to \cite{xu_journeyBP}). One \ac{PE} updates the $\mathbf{L}$- and $\mathbf{R}$-messages as follows \cite{xu_journeyBP}:
\begin{equation}\label{bp-update}
\begin{array}{cc}
L_{out,1}= & f\left(L_{in,1},L_{in,2}+R_{in,2}\right)\\
R_{out,1}= & f\left(R_{in,1},L_{in,2}+R_{in,2}\right)\\
L_{out,2}= & f\left(R_{in,1},L_{in,1}\right)+L_{in,2}\\
R_{out,2}= & f\left(R_{in,1},L_{in,1}\right)+R_{in,2}
\end{array}
\end{equation}
where $$f\left(a,b\right)=\mathrm{ln}\left(\frac{1+e^{a+b}}{e^{a}+e^{b}}\right).$$
For initialization, all messages are set to zero, except for the first and last stage, where
\begin{equation}
\begin{split}
{L}_{i,n+1} & ={L}_{i,ch} \textrm{, and} \\
{R}_{i,1} &=
   \begin{cases}
      {L}_{max} & \text{} \forall i \in \mathbb{\bar{A}}\\
      0 & \text{else}\\
   \end{cases}
\end{split}
\end{equation}
with $\mathit{L}_{max}$ denoting the clipping value of the decoder (in theory: $L_{max}\to\infty$), as all values within the simulation are clipped to be within $(-\mathit{L}_{max},\mathit{L}_{max})$. This prevents from experiencing numerical instabilities.

\subsection{Partitionable Codes}

As opposed to other “random-like” channel codes with close-to-capacity performance, polar codes exhibit a very regular (algebraic) structure.
It is instructive to realize that the encoding graph, as visualized in Fig.~\ref{encoding-circuit} for $N=8$, can be partitioned into independent sub-graphs \cite{hashemi_partitioned}, \cite{tse_parallelDecoder}, i.e., there is no interconnection in the first $\mathrm{log_2}(N_p)$ stages, where $N_p$ denotes the number of bits per sub-block. We define a \emph{partitionable} code in a sense that each sub-block can be decoded independently (i.e., no interconnections within the same stage exist; leading to a tree like factor graph). This algorithm can be adopted to all partitionable codes and is not necessarily limited to polar codes.
Each sub-graph (in the following called sub-block) is now coupled with the other sub-blocks only via the remaining polar stages as depicted in Fig.~\ref{encoding-circuit}.
In order to simplify polar decoding, several sub-blocks $B_i$ can now be decoded on a per-sub-block basis \cite{hashemi_partitioned}. The set of frozen bit positions $\mathbb{\bar{A}}$ need to be split into sub-sets $\mathbb{\bar{A}}_i$ corresponding to the sub-blocks $B_i$ (with information vector $\mathbf{u}_i$) and, thus, each sub-block might show a different code rate.

In a more abstract view, we follow the spirit of \cite{Sarkis_SSC}. Their \ac{SSC} algorithm partitions the decoding tree into \ac{SPC}  and \ac{RC}. This turns out to be a very efficient way of improving the overall throughput, as the \ac{SPC} and {RC} sub-decoders can be efficiently implemented. Our approach replaces these partitions by \ac{NN}s.

\subsection{Deep-Learning for Channel Coding}\label{deep-learning}

For the fundamentals of deep learning, we refer the reader to \cite{goodfellow2016}. However, for the sake of terminology, we provide a brief overview on the topic of machine learning. For a more detailed explanation how to train the sub-block \ac{NND} we refer to \cite{gruber_CISS}. Feedforward \acp{NN} consist of many connected neurons which are arranged in $L$ layers without feedback connections. The output $y$ of each neuron depends on its weighted inputs $\theta_i x_i$ and its activation function $\operatorname{g}$, given by 
\begin{align}
y = \operatorname{g} \left( \sum\limits_i \theta_i x_i +\theta_0\right).
\end{align}
The whole network composes together many different functions $\mathbf{f}^{\left(l\right)}$ of each layer $l$ and describes an input-output mapping 
\begin{align}
 \mathbf{w} =  \mathbf{f}\left( \mathbf{v}; \boldsymbol{\Theta} \right)  = \mathbf{f}^{\left(L-1 \right)} \left( \mathbf{f}^{\left(L-2 \right)} \left( \ldots \left( \mathbf{f}^{\left(0 \right)} \left( \mathbf{v} \right)  \right) \right) \right)    
\end{align}
where $\mathbf{v}$, $\mathbf{w}$ and $\boldsymbol{\Theta}$ denote the input vector, output vector and the weights of the \ac{NN}, respectively.  It was shown in \cite{hornik1989} that such a multi-layer \ac{NN} with $L=2$ and nonlinear activation functions can theoretically approximate any continuous function on a bounded region arbitrarily closely---if the number of neurons is large enough. A training set of known input-output mappings is required in order to find the weights $\boldsymbol{\Theta}$ of the \ac{NN} with gradient descent optimization methods and the backpropagation algorithm \cite{rumelhart1986}. After training, the \ac{NN} is able to find the right output even for unknown inputs which is called \emph{generalization}.

As described in \cite{gruber_CISS}, we use a feed-forward deep \ac{NN} that can learn to map a noisy version of the codeword to its corresponding information bits. Each hidden layer employs \ac{ReLU} activation functions and the final stage is realized with a sigmoid activation function \cite{goodfellow2016} in order to obtain output values in between zero and one giving the probability that the output bit is ``1''. In order to keep the training set small, we extend the decoder \ac{NN} with additional layers which model an abstract channel \cite{gruber_CISS}, i.e., a training set containing every codeword is sufficient to train with as much training samples as desired. 

\begin{figure}[ht]
\vspace*{0.3cm}
 \centering
    \begin{tikzpicture}
  \begin{axis}[
            width=\linewidth,
            height=0.225\textheight,
            xmajorgrids,
            yminorticks=true,
            ymajorgrids,
            yminorgrids,
            legend pos=north west,
            legend entries={$N=32$, $N=64$},
            legend style={legend cell align=left,align=left,draw=white!15!black, font=\footnotesize},
            xlabel={\# information bits $k$},
            ylabel={gap to MAP perf.  $\left[ \text{dB} \right]$},
            xmin=8,
            xmax=14,
            ymin=0,
            ymax=0.8
        ]
        \addlegendimage{color=apfelgruen, mark=*, thick}
	\addlegendimage{color=mittelblau, mark=diamond*, thick}

        \addplot[color=apfelgruen, mark=*, thick] table {data/gap2mapN=32.data};
        \addplot[color=mittelblau, mark=diamond*, thick] table {data/gap2mapN=64.data};

  \end{axis}
\end{tikzpicture}
 \caption{Scalability shown by the gap to \ac{MAP} performance in dB at \ac{BER} = 0.01 for \unit[32/64]{bit}-length polar codes with different code rates for fixed learning epoches.  }
 \label{fig:dimensionality} 
\end{figure}
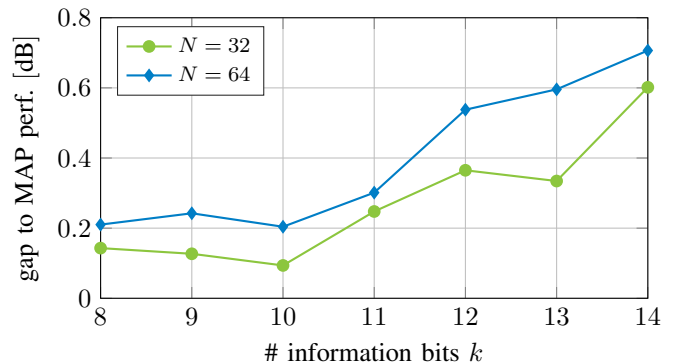

It was shown in \cite{gruber_CISS}, that it is possible to decode polar codes with \ac{MAP} performance for small block lengths. The \ac{BER} performance gap between a \ac{NND} and \ac{MAP} decoding is shown in Fig. \ref{fig:dimensionality}. It illustrates that learning to decode is limited through exponential complexity as the number of information bits in the codewords increases.

\section{Partitioned Polar Neural Network Decoding}

Instead of using sequential \ac{SCL} sub-block decoders as in \cite{hashemi_partitioned}, we replace them by \ac{NN} decoders. Every sub-block $\mathrm{NN}_i$ covers $k_i$ information bits.  As the number of possible encoder output states (i.e., different codewords) is $2^{k_i}$, efficient training is only possible for small sub-blocks, containing a small amount of information bits $k_i$ \cite{gruber_CISS}.
The advantage of this concept is that each sub-block \ac{NN} decoder can be trained independently. However, each \ac{NN} has its own corresponding frozen bit positions $\mathbb{A}_i$ and a specific block lengths $N_i$. As we deal with man made signals, the training and validation dataset can be created by random input data, a conventional polar encoder and a channel. Thus, an infinite amount of labeled training data is available.

Now, the \ac{NN} decoder can be efficiently trained offline to \ac{MAP} performance \cite{gruber_CISS}, since the effective block-size per sub-block $N_i$ reduces by the number of sub-blocks $M$.
Each \ac{NN} decoder outputs the decoded codeword (and/or the extracted information bits) either as soft-values, i.e., probabilities, or after a hard-decision which might require re-encoding.  These bits are now treated as known values.

\begin{center}
\begin{figure}
\vspace*{0.3cm}
\includegraphics[width=\columnwidth]{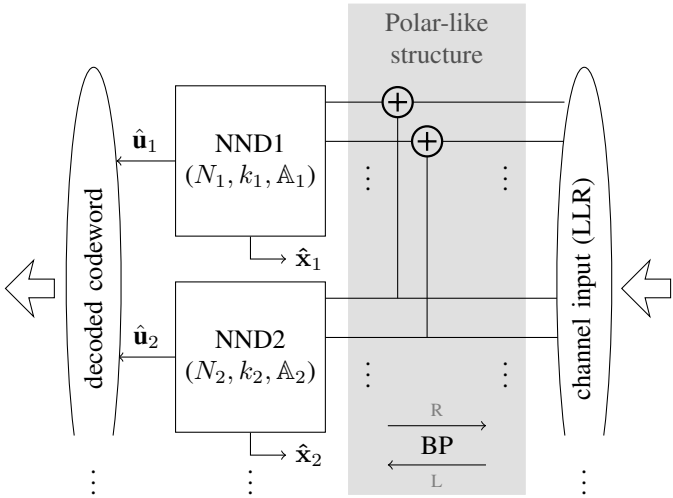}
\caption{Partitioned neural network polar decoding structure.}
\label{fig:general-structure}
\end{figure}
\end{center}

Our proposed decoder consists of two stages (see Fig. \ref{encoding-circuit} and Fig. \ref{fig:general-structure}): 
\begin{enumerate}
\item $M$ deep learning blocks, trained to decode the corresponding sub-codeword with length $N_i$ and frozen bit position vector $\mathbb{A}_i$
\item A \emph{conventional} \ac{BP} part, when the already decoded sub-blocks are propagated via the \emph{coupling} stages of the remaining polar encoding graph.
\end{enumerate}

After initializing the rightmost stages with the received channel \ac{LLR}-values, messages are propagated from stage to stage according to the \ac{BP} update rules in (\ref{bp-update}) until the first \ac{NN} decoder stage $n_{\text{NN}}$ (see Fig. \ref{encoding-circuit}). Then, the first \ac{NN} decoder estimates the received sub-block $\mathrm{NN}_1$.
After having decoded the first sub-block, the results are propagated via conventional \ac{BP} decoding algorithms through the remaining \emph{coupling} stages (see Fig. \ref{encoding-circuit} and \ref{fig:general-structure}). Thus, this algorithm is block-wise sequential where $M$ sub-blocks $\mathrm{NN}_i$ are sequentially decoded from top to bottom.
The detailed decoding process works as follows:
\begin{enumerate}
\item Initialize stage $n+1$ with received channel \ac{LLR}-values 
\item Update stages $n_{\text{NN}}$ to $n+1$ according to \ac{BP} algorithm rules (\emph{propagate \acp{LLR}})
\item Decode next sub-block (top to bottom)
\item Re-encode results and treat results as perfectly known, i.e., as frozen bits
\item If not all sub-blocks are decoded go to step 2
\end{enumerate}

The interface between the \ac{NND} and the BP stages depends on the trained input format of the \ac{NN}. Typically, the \ac{NN} prefers normalized input values between 0 and 1. 
Fortunately, it was observed in \cite{gruber_CISS} that the \ac{NN} can handle both input formats and effective training is possible.

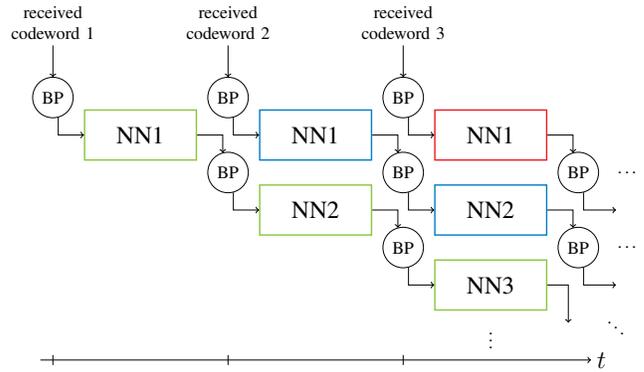
\begin{figure}
  \centering
  \resizebox{\linewidth}{!}{

  \begin{tikzpicture}
	  \tikzstyle{block} = [rectangle, draw,  text width=14em, minimum height=1cm, text width=2cm, align=center];
  \tikzstyle{dots}=[circle, inner sep=0.5pt, fill=black]
      \node[draw,circle] (bp0a) at (-1.75,0.75) {BP};
      \node[block,draw=apfelgruen,thick] (nn1a) at (0,0) {\Large NN1};
      \node[draw,circle] (bp1a) at (1.75,-0.75) {BP};
      \node[block,draw=apfelgruen,thick] (nn2a) at (3.5,-1.5) {\Large NN2};
      \node[draw,circle] (bp2a) at (5.25,-2.25) {BP};
      \node[block,draw=apfelgruen,thick] (nn3a) at (7,-3) {\Large NN3};
      \coordinate (v1) at (8.75,-3.75);
      \node[text width=2cm,align=center] (y1) at (-1.75,2.25) {received codeword 1};

      \draw[->] (y1) -- (bp0a);
	\draw[->] ([xshift=0.2cm]bp0a) |- (nn1a);
    \draw[->] (nn1a) -| ([xshift=-0.2cm]bp1a);
    \draw[->] ([xshift=0.2cm]bp1a) |- (nn2a);
    \draw[->] (nn2a) -| ([xshift=-0.2cm]bp2a);
    \draw[->] ([xshift=0.2cm]bp2a) |- (nn3a);
    \draw[->] (nn3a) -| ([xshift=-0.2cm]v1);

     \node[draw,circle] (bp0b) at (1.75,0.75) {BP};
      \node[block,draw=mittelblau,thick] (nn1b) at (3.5,0) {\Large NN1};
      \node[draw,circle] (bp1b) at (5.25,-0.75) {BP};
      \node[block,draw=mittelblau,thick] (nn2b) at (7,-1.5) {\Large NN2};
      \node[draw,circle] (bp2b) at (8.75,-2.25) {BP};
 \coordinate (v2) at (12.25,-3.75);
 \node[text width=2cm,align=center] (y2) at (1.75,2.25) {received codeword 2};

\draw[->] (y2) -- (bp0b);
	\draw[->] ([xshift=0.2cm]bp0b) |- (nn1b);
    \draw[->] (nn1b) -| ([xshift=-0.2cm]bp1b);
    \draw[->] ([xshift=+0.2cm]bp1b) |- (nn2b);
    \draw[->] (nn2b) -| ([xshift=-0.2cm]bp2b);

    \node[draw,circle] (bp0c) at (5.25,0.75) {BP};
    \node[block,draw=rot,thick] (nn1c) at (7,0) {\Large NN1};
      \node[draw,circle] (bp1c) at (8.75,-0.75) {BP};
    \coordinate (v3) at (9.5,-3);
\coordinate (v4) at (9.5,-1.5);
 \node[text width=2cm,align=center] (y3) at (5.25,2.25) {received codeword 3};
\draw[->] (y3) -- (bp0c);
	\draw[->] ([xshift=0.2cm]bp0c) |- (nn1c);
    \draw[->] (nn1c) -| ([xshift=-0.2cm]bp1c);
    \draw[->] ([xshift=0.2cm]bp2b) |- (v3);
 \draw[->] ([xshift=0.2cm]bp1c) |- (v4);

   \node at (9.75,-0.75) {\ldots};
\node at (9.75,-2.25) {\ldots};
 \node at (7,-4) {\vdots};
 \node at (9.5,-3.75) {$\ddots$};




 \draw[->] (-2,-4.5) -- (9,-4.5);
\node[anchor=west] at (9,-4.5) {\Large $t$};

\foreach \x in {-1.75,1.75,5.25} {
  \draw (\x,-4.6) -- (\x,-4.4);
} 


  \end{tikzpicture}
  }
  \caption{Pipelined implementation of the proposed PNN decoder.}
  \label{fig:decoding-pipeline}
\end{figure}

In summary, the system itself can be modeled as one large \ac{NN} as well. Each \ac{BP} update iterations defines additional layers, which are deterministic and thus do not effect the training complexity, similar to regularization layers \cite{goodfellow2016}. This finally leads to a pipelined structure as depicted in Fig. \ref{fig:decoding-pipeline}. 
As each \ac{NN} is only passed \emph{once} and to emphasize the difference compared to iterative decoding, we term this kind of decoding as \textit{one-shot-decoding}.

\subsection{Further Optimizations}

As it can be seen in Fig. \ref{fig:dimensionality}, the limiting parameter is the number of information bits per sub-block $k_i$, since it defines the number of possible estimates of the \ac{NND}.
One further improvement in terms of sub-block size can be done by merging multiple equally-sized sub-blocks such that $\sum_i k_i<k_{\text{max}}$, leading to an unequal sub-block size. This helps to obtain as few as possible sub-blocks. The results for $M=8$ partitions are shown in Fig. \ref{fig:ber_sim}.

Additionally, \emph{finetuning-}learning can be applied to the overall network in order to adjust the independently trained components to the overall system. This means that the whole decoding setup is used to re-train the system such that the conventional stages are taken into consideration for decoding. This prevents from performance degradation due to potentially non-Gaussian NN-input distributions, which was assumed during the training. Such an effect can be observed whenever clipping of the LLRs is involved. However, the basic structure is already fixed and thus only a small amount of the $2^k$ possible codewords is sufficient for good training results. The required training set is created with the free-running decoder.

The coupling could be also done by an \ac{SC} stage (as originally  proposed in \cite{hashemi_partitioned}) without having additional iterations. However, the BP structure suits better to the \ac{NN} structure as both algorithms can be efficiently described by a graph and their corresponding edge weights. Thus, the BP algorithm is preferred.

For \ac{CRC} aided decoding (a \ac{CRC} check over the whole codeword) the \ac{CRC} check can be split into smaller parts as in \cite{hashemi_partitioned}, where each \ac{CRC} only protects one sub-block, i.e., the \ac{CRC} can be considered by the \ac{NN} decoders and thus straightforwardly learned. However, this requires at least some larger NN-sub-blocks, otherwise the rate-loss due to the CRC checks becomes prohibitive and is thus not considered at the moment.

\section{Comparison with SCL/BP}

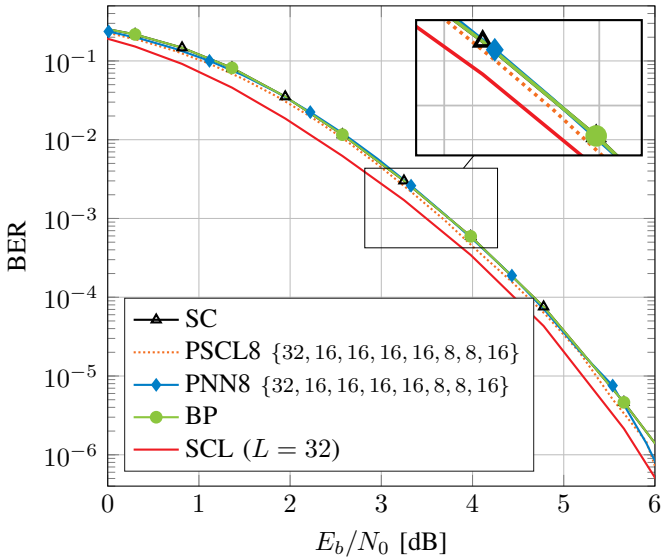
\begin{figure}[t]
\begin{tikzpicture}[spy using outlines={rectangle, magnification=1.7, size=0.5cm, connect spies}]
    \begin{axis}[
        width=\columnwidth,
        height=0.9\columnwidth,
        xmajorgrids,
        ymode=log,
        yminorticks=true,
        ymajorgrids,
        legend pos=south west,
        legend style={legend cell align=left,align=left,draw=white!15!black},
        xlabel={$E_b/N_0$ [dB]},
        ylabel={BER},
        xmin=0,    
        xmax=6,
        ymin=0.0000004,
        ymax=0.5        
    ]

    \addplot[black,thick,mark=triangle,mark repeat=1.5] table {data/128bit_sc_A_fabian.dat};
    \addlegendentry{SC};
    
    \addplot[orange,thick, mark repeat=3,densely dotted] table {data/128bit_pscl_A_fabian.data};
    \addlegendentry{PSCL8 \footnotesize{$\{32,16,16,16,16,8,8,16\}$}};
    
    \addplot[solid,mittelblau,thick, mark=diamond*,mark repeat=3] table {data/128bit_pnn_A_fabian.data};
    \addlegendentry{PNN8 \footnotesize{$\{32,16,16,16,16,8,8,16\}$}};

    \addplot[solid,apfelgruen,thick,mark=*,mark repeat=2] table {data/128bit_bp_A_fabian.data};
    \addlegendentry{BP};
    
    \addplot[solid,rot,thick] table {data/128bit_scl_A_fabian.data};
    \addlegendentry{SCL ($L=32$)};
    


    \begin{scope}
    	\spy[black,width=3cm,height=1.8cm] on (4.3,3.7) in node [fill=none] at (5.6,5.3);
    \end{scope}
    
    \end{axis}
\end{tikzpicture}
\caption{\ac{BER} performance of the proposed partitioned NN decoder for $N=128$ and $M=8$ partitions with variable size in comparison to state-of-the-art SC, BP and SCL ($L=32$) decoding.}
\label{fig:ber_sim}
\end{figure}

In general, a fair comparison with existing solutions is hard, as many possible optimizations need to be considered. We see this idea as an alternative approach, for instance in cases whenever low-latency is required. The \ac{BER} results for $N=128$ and different decoding algorithms are shown in Fig.~\ref{fig:ber_sim}. As shown in Tab \ref{tab:subblocks}, the size of the partitions is chosen such each partition does not contain more than $k_{\text{max}} = 12$ information bits, which facilitates learning the sub-blocks. If the sub-block does not contain any frozen bits, a simple hard-decision decoder is used. Basically, the concept of partitioning polar codes helps to scale \ac{NND} to longer codes. 

\begin{table}
\vspace*{0.3cm}
 \caption{Number of information bits $k_i$ for each sub-block}
 \begin{tabular}{lcccccccc}
  sub-block $i$ & 1 & 2 & 3 & 4 & 5 & 6 & 7 & 8 \\ \hline
  sub-block size $N_i$ & 32 & 16 & 16 & 16 & 16 & 8 & 8 & 16 \\
  information bits $k_i$ & 1 & 3 & 11 & 5 & 13 & 7 & 8 & 16 \\
 \end{tabular}
 \label{tab:subblocks}
\end{table}

We denote a \ac{PNN} with $M$ partitions as PNN$M$. Although each \ac{NN} in the \ac{PNN} can be learned to approximately \ac{MAP} performance, there is a loss $\lambda$ between \ac{PNN} decoding and conventional \ac{SCL} decoding with list size $L$. We limit ourselves in this work to $L=32$. The loss $\lambda$ can be explained by two main reasons:
\begin{enumerate}
\item Loss through partitioning $\lambda_{\text{part}}$ as the concept only applies sub-block \ac{MAP} decoding.
\item Loss due to sub-optimality of the \ac{NND} $\lambda_{\text{NN}}$, due to insufficient training or non-Gaussian input distributions.\footnote{We did not focus on how to find the best \ac{NN} structure for each \ac{NN} because we want to introduce the concept of partitioned polar codes in order to scale \ac{NND}. We expect that for sufficient training and hyperparameter tuning \cite{bergstra2011} this loss vanishes.}
\end{enumerate}

For further analysis of this problem, we replace the \acp{NND} in each partition by \ac{SCL} decoders with list size $L=32$ as in \cite{hashemi_partitioned} and obtain a \ac{PSCL} decoder. As before, PSCL$M$ terms a \ac{PSCL} decoder with $M$ partitions. This enables the investigation of larger partition sizes and thus the effect of partitioning, which is currently not possible with \acp{NN} due to the limited training length. Since \ac{SCL} decoding approximates \ac{MAP} performance for properly chosen parameters \cite{talvardyList}, the \ac{PSCL} gives a lower bound on the expected \ac{BER}  according to the actual number of partitions. This enables to observe $\lambda_{\text{part}}$ and $\lambda_{\text{NN}}$ separately. 

In order to quantify the loss, we introduce a similar concept as in \cite{gruber_CISS}: the \ac{NE} for a decoding concept, e.g., \ac{PNN}, \ac{PSCL}, which is given by
\begin{align}
 \text{NE} = \frac{1}{D} \sum\limits_{t=1}^D \frac{\text{BER} \left( \rho_t \right) }{\text{BER}_{\text{SCL}} \left( \rho_t \right)}
\end{align}
where $\rho_t$ is a certain \ac{SNR} value, $\text{BER} \left( \rho_t \right)$ denotes the decoders \ac{BER} for this specific \ac{SNR} and $\text{BER}_{\text{SCL}} \left( \rho_t \right)$ is the corresponding \ac{BER} of the \ac{SCL} decoder, respectively. Thus, NE compares the decoding performance over a range of $D$ different SNR values $\rho_1, \ldots, \rho_T$.

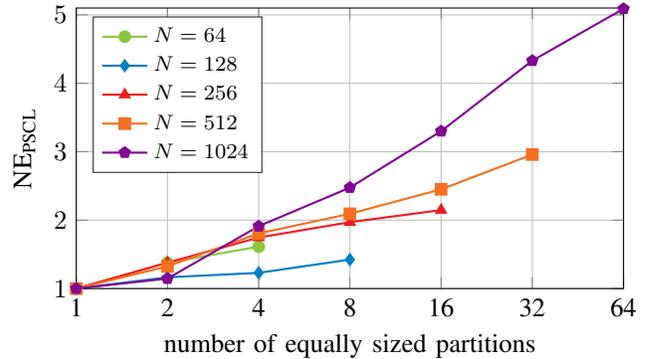
\begin{figure}[t]
\vspace*{0.3cm}
\begin{tikzpicture}
  \begin{axis}[
            width=\linewidth,
            height=0.225\textheight,
            xmajorgrids,
            yminorticks=true,
            ymajorgrids,
            yminorgrids,
            legend pos=north west,
            legend entries={$N=64$, $N=128$, $N=256$, $N=512$, $N=1024$},
            legend style={legend cell align=left,align=left,draw=white!15!black, font=\footnotesize},
            xlabel={number of equally sized partitions},
            ylabel={${\text{NE}}_{\text{PSCL}}$},
            xmin=1,
            xmax=64,
            ymin=1,
            ymax=5.1,
	    xmode=log,
	    xtick = {1,2,4,8,16,32,64},
	    xticklabels = {1,2,4,8,16,32,64}
        ]
        \addlegendimage{color=apfelgruen, mark=*, thick}
	\addlegendimage{color=mittelblau, mark=diamond*, thick}
	\addlegendimage{color=rot, mark=triangle*, thick}
	\addlegendimage{color=orange, mark=square*, thick}
	\addlegendimage{color=lila, mark=pentagon*, thick}
        
        \addplot[color=apfelgruen, mark=*, thick] table {data/normalized_error_partitionSize64.dat};
        \addplot[color=mittelblau, mark=diamond*, thick] table {data/normalized_error_partitionSize128.dat};
        \addplot[color=rot, mark=triangle*, thick] table {data/normalized_error_partitionSize256.dat};
        \addplot[color=orange, mark=square*, thick] table {data/normalized_error_partitionSize512.dat};
        \addplot[color=lila, mark=pentagon*, thick] table {data/normalized_error_partitionSize1024.dat};

  \end{axis}
\end{tikzpicture}
\caption{Effect of the partitioning on the \ac{BER} performance. \disable{\sebastian{***change x-axis into number of equally sized partitions***}}}
\label{fig:nve_comparison}
\end{figure}

\begin{figure}[t]
 \includegraphics[width=\columnwidth]{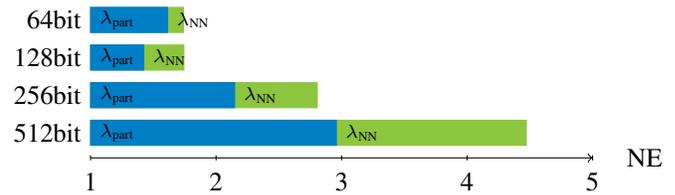}
\caption{Normalized error ${\text{NE}}_{\text{PSCL}}$ due to $\lambda_{\text{part}}$ (blue) and ${\text{NE}}_{\text{PNN}}$ due to $\lambda_{\text{NN}}$ (green) for fixed size  of partitions $N_i=16$.}
\label{fig:error_bar}
\end{figure}

It can be observed in Fig.~\ref{fig:nve_comparison}, that $\text{\ac{NE}}_{\text{PSCL}}$ increases with the number of partitions, i.e.,  decreases with the partition sizes. Fig.~\ref{fig:error_bar} relates the effect of $\lambda_{\text{part}}$ and  $\lambda_{\text{NN}}$. It can be observed that the main error originates from partitioning and only a small part from suboptimal \acp{NN}. The amount of sub-blocks with larger $k_i$ increases with larger codelength and at the same time it is more difficult to achieve \ac{MAP} performance for these blocks \cite{gruber_CISS}. Therefore, the loss $\lambda_{\text{NN}}$ becomes more important for longer codes. The training complexity limits the feasible sub-block size of \ac{PNN} decoding and thus long codes require a lot of partitions. The larger the number of partitions, the larger $\lambda_{\text{part}}$.  However, the fast progress in the machine learning domain might enable larger sub-blocks, which should improve  $\lambda_{\text{part}}$ and therefore the overall performance of the \ac{PNN} concept.

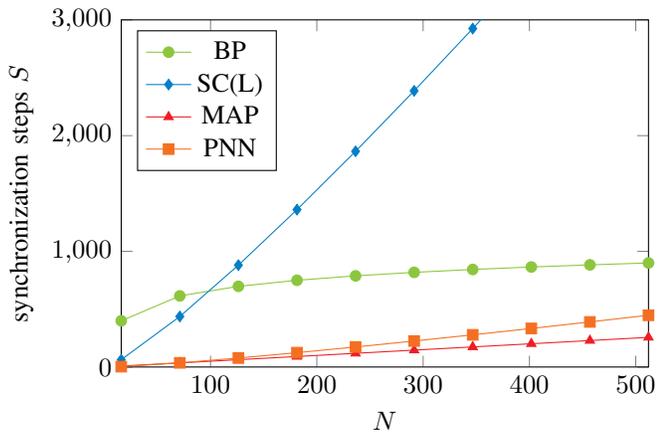
\begin{figure} 
  \def\iter{50}
\def\L{32}
\def\P{16}
\def\D{30}
\def\h{3}
\def\r{0.5}

\begin{tikzpicture}
\begin{axis}[width=0.97\columnwidth,
        height=0.7\columnwidth,
domain=16:512,
                       samples=10,
		       xlabel={$N$},
                       ylabel={synchronization steps $S$},
                       ymin=1,
		       ymax=3000,
xmin=16,
xmax=512,
legend pos=north west]
      
        \addplot[color=apfelgruen,mark=*] {2*\iter*log2(x)};
        \addlegendentry{BP};
        
        \addplot[color=mittelblau,mark=diamond*]  {x*log2(x)};
        \addlegendentry{SC(L)};

	\addplot[color=rot,mark=triangle*] {1+\r*x};
        \addlegendentry{MAP};

        \addplot[color=orange,mark=square*] {x/\P*(\h+1) + x/\P*2*log2(x/\P) };
        \addlegendentry{PNN};
    \end{axis}
\end{tikzpicture}
  \label{fig:complexity_syncs}
\caption{Comparison of required synchronization steps for different polar decoding approaches.}
\label{fig:complexity}

\end{figure}

In order to approximate the decoding latency in Fig.~\ref{fig:complexity}, we count the number of synchronization steps as operations can be done in parallel, but need to be synchronized after each synchronization step. The latency of the \ac{SCL} algorithm can be described by $\mathcal{O} \left( N \log N \right)$ because it estimates each bit sequentially. Thus, the number of required synchronization steps is $$S_{\text{SCL}}=N \log N.$$ The \ac{BP} algorithm scales much better with $\mathcal{O} \left( \log N \right)$ because synchronization is required after each \ac{BP} stage but depends on the number of iterations $I$, namely, $$S_{\text{BP}}=2I \log N.$$ The \ac{PNN} decoder enforces less \ac{BP} updates and the \acp{NND} itself only synchronize after each layer $$S_{\text{PNN}}= \frac{N}{N_P} N_H + \frac{N}{N_P} 2\log \frac{N}{N_P}$$ where $N_P$ and $N_H$ respectively denote the size of each partition and the number of hidden layers in the \ac{NN}. Fig.~\ref{fig:complexity} is given for $N_P=16$ and $N_H=3$.
To sum up, our approach enables latency reduction of \ac{BP} decoding, while being competitive with the \ac{SC} and \ac{BP} \ac{BER} performance.

\section{Conclusion}

In this work, we have shown that one way to reach scalability of deep-learning based channel decoding can be described by replacing sub-components of an existing decoder by \ac{NN}-based components.
This enables scalability in terms of block length and number of information bits towards practical lengths. Meanwhile, the length is still limited to short codes as the degradation through partitioning limits the overall performance of this concept. The \ac{BER} performance of our decoder turns out to be similar to the \ac{SC} and \ac{BP} performance. However, the latency reduces a lot when the inherent parallel structure of this algorithm is exploited, since \emph{one-shot-decoding} (i.e., non-iterative decoding) becomes possible. We have shown that the performance degradation is mainly a result of the small partitions, as the sub-block size is currently strictly limited by the training. Nonetheless, the proposed setup would scale very well for larger sub-blocks, therefore future work needs to be done on potential improvements of the NN structure such that larger components become available.

\bibliographystyle{IEEEtran}
\bibliography{references}

\end{document}